\documentclass[epj]{svjour}
\usepackage{graphics}
\begin{document}
\title{Rotating Black Branes in the presence of nonlinear electromagnetic
field}
\author{S. H. Hendi
\thanks{\emph{Present address:} hendi@mail.yu.ac.ir}%
}                     
\institute{Physics Department, College of Sciences, Yasouj
University, Yasouj 75914, Iran \and Research Institute for
Astrophysics and Astronomy of Maragha (RIAAM), P.O. Box 55134-441,
Maragha, Iran \and National Elite Foundation, P.O. Box 19615-334,
Tehran, Iran}
\date{Received: date / Revised version: date}
%
\abstract{ In this paper, we consider a class of gravity whose
action represents itself as a sum of the usual Einstein-Hilbert
action with cosmological constant and an $U(1)$ gauge field for
which the action is given by a power of the Maxwell invariant. We
present a class of the rotating black branes with Ricci flat
horizon and show that the presented solutions may be interpreted
as black brane solutions with two event horizons, extreme black
hole and naked singularity provided the parameters of the
solutions are chosen suitably. We investigate the properties of
the solutions and find that for the special values of the
nonlinear parameter, the solutions are not asymptotically
anti-deSitter. At last, we obtain the conserved quantities of the
rotating black branes and find that the nonlinear source effects
on the electric field, the behavior of spacetime, type of
singularity and other quantities.
\PACS{
      {PACS-key}{discribing text of that key}   \and
      {PACS-key}{discribing text of that key}
     } 
} 
\maketitle
\section{Introduction}
\label{intro}Seventy five years ago Born and Infeld proposed a
type of nonlinear electrodynamics starting from the principle of
finiteness with the aim of obtaining a finite value for the
self-energy of a point-like charge \cite{BI1,BI2}. They proposed
the following Lagrangian
\begin{equation}
L(\mathcal{F})=4\beta ^{2}\left( 1-\sqrt{1+\frac{\mathcal{F}}{2\beta ^{2}}}%
\right) .
\end{equation}%
where the Maxwell invariant $\mathcal{F}=F_{\mu \nu }F^{\mu \nu
}$\ in which $F_{\mu \nu }=\partial _{\mu }A_{\nu }-\partial _{\nu
}A_{\mu }$ is the electromagnetic tensor field and $A_{\mu }$ is
the vector potential. After these significant achievements, there
were not many works on nonlinear electromagnetic fields, as
Plebanski mentioned in $1970$ at the beginning of his monograph
\cite{Plebanski}: \textquotedblleft \textit{If in recent times the
interest in nonlinear electrodynamics cannot be said to be very
popular, it is not due to the fact that one could rise some
serious objections against this theory. It is simply rather
difficult in its mathematical formulation, what causes that it is
very unlikely to derive some concrete results in closed
form}\textquotedblright . The major responsible of the actual
revival, more than twenty five years later of the moment of
formulation of quoted statement, is the fact that the effective
action for
the open string ending on $D$-branes can be written in a nonlinear form \cite%
{FradTsey,Tsey,SeibWitt}. In recent years nonlinear
electrodynamics models are attracting much interest, too. The
nonlinear equations of theoretical and mathematical physics
attract the significant attention because of the specific
properties such as the absence of the superposition principle, the
nonlinear fields interactions and existence of the soliton
solutions. Some of the main reasons to consider nonlinear
electrodynamics theories are as follows: I first point out that
these theories are considerably richer than that of the classical
linear electromagnetic field and in the special case they can
reduce to Maxwell field. Second, it can be used to find and
analyze new solutions with different behavior of the nonlinear
Maxwell equations. Third, it is remarkable that all the nonlinear
electrodynamics models coupled to gravity satisfy the zeroth and
the first laws of black hole mechanics \cite{Rasheed}. Fourth,
there has been a significant amount of interest in cosmological
models involving nonlinear electromagnetic fields
\cite%
{Novello11,Novello12,Novello13,Novello14,Novello2,Camara1,Camara2,Camara3,Camara4,Camara5,Camara6,Camara7,Camara8,Camara9,Camara10,Camara11,Camara12,Camara13,Camara14,Camara15}%
. If the early universe is dominated by radiation governed by
Maxwell's equations it is well known that there will be a
spacelike initial singularity in the past. However, if Maxwell's
equations become modified in the early universe, when the
electromagnetic field is large, it may be
possible to avoid the initial singularity. In fact, recent results \cite%
{Novello11,Novello12,Novello13,Novello14} show that a magnetic
universe can avoid the initial singularity and have a period of
late time acceleration if the electromagnetic Lagrangian is of the
nonlinear form \cite{Vollick}. More recently, in the light of the
AdS/CFT correspondence \cite{AhGubMald}, the nonlinear
electrodynamics string approach has been used to obtain solutions
describing baryon configurations which are consistent with confinement \cite%
{Callan}. Moreover, in \cite{Vargas1,Vargas2} it was found that
cosmological models with nonlinear electromagnetic fields can be
compatible with the recently measured accelerated expansion of the
universe.

From this point of view, the exact solutions of Einstein-nonlinear
Maxwell equations are worth to study since they may indicate the
physical relevance of nonlinear effects in strong gravitational
and strong electromagnetic fields in cosmological and general
relativity models. In nonlinear electrodynamics coupled to general
relativity there exists several solutions
describing electrically charged black holes and black branes \cite%
{Garcia1,Garcia2,Garcia3,Garcia4,Breton,BHnon1,BHnon2,BHnon3,BHnon4,BHnon5,BHnon6,BHnon7,BHnon8,BHnonmartinez1,BHnonmartinez2,BHnonmartinez3,BHnonmartinez4,BlackBrane1,BlackBrane2,BlackBrane3}%
, magnetic black holes and magnetic branes \cite%
{Bronnikov,MagBrane1,MagBrane2}. Also, the solutions of the higher
derivative gravity coupled to nonlinear electromagnetic fields
have been
well studied in the literature \cite{Higher1,Higher2,Higher3,Higher4,Higher5}%
.

The aim of this paper is to consider the nonlinear electrodynamics
field coupled to Einstein gravity and introduce a class of
rotating black brane solutions with flat horizon. Also, we present
a study of the effects produced by a nonlinear type field on the
asymptotic behavior of the solutions and conserved quantities. The
plan of the paper is as follows: We give a brief review of the
finite well-defined action and field equations of
Einstein gravity in the presence of nonlinear electromagnetic field in Sec. %
\ref{Fiel}. Then, in Sec. \ref{Rotating Sol}, we present a new
class of rotating black brane solutions with flat horizon in
Einstein-nonlinear Maxwell gravity, investigate their properties
and obtain conserved quantities of the $(n+1)$-dimensional black
brane solutions. We finish our paper with some concluding remarks.

\section{Basic Field Equations of Einstein Gravity with nonlinear
Electromagnetic Source}

\label{Fiel}

The $(n+1)$-dimensional action in which gravity is coupled to
nonlinear electrodynamics field is given by
\begin{equation}
\mathcal{I}=-\frac{1}{16\pi }\int\limits_{\mathcal{M}}d^{n+1}x\sqrt{-g}%
\left( \mathcal{L}_{1}+\mathcal{L}_{2}\right) ,  \label{Action}
\end{equation}%
where $\mathcal{L}_{1}=\mathcal{R}-2\Lambda $,
$\mathcal{L}_{2}=-\alpha \mathcal{F}^{s}$, $\mathcal{R}$ is scalar
curvature, $\Lambda $ refers to
the negative cosmological constant which is in general equal to $%
-n(n-1)/2l^{2}$ for asymptotically AdS solutions, in which $l$\ is
a scale length factor, $\alpha $ denotes a coupling constant and
the exponent $s$ is related to the power of nonlinearity. Varying
the action with respect to the metric $g_{\mu \nu }$ and the gauge
field $A_{\mu }$, the field equations are obtained as
\begin{equation}
R_{\mu \nu }-\frac{1}{2}g_{\mu \nu
}\mathcal{L}_{1}=\mathrm{T}_{\mu \nu }, \label{GravEq}
\end{equation}%
\begin{equation}
\partial _{\mu }\left( \sqrt{-g}F^{\mu \nu }\mathcal{F}^{s-1}\right) =0,
\label{MaxEq}
\end{equation}%
where
\begin{equation}
\mathrm{T}_{\mu \nu }=2\alpha \mathcal{F}^{s-1}\left( sF_{\mu \rho
}F_{\nu }^{\rho }-\frac{1}{4}g_{\mu \nu }\mathcal{F}\right) .
\label{TNonMax}
\end{equation}%
For a well-defined variational principle \cite{Myers1,Myers2}, one
has to supplement the action (\ref{Action}) with the
Gibbons-Hawking boundary term
\begin{equation}
\mathcal{I}_{b}=-\frac{1}{8\pi }\int_{\partial \mathcal{M}}d^{n}x\sqrt{%
-\gamma }K,  \label{Ib}
\end{equation}%
where $\gamma $ and $K$ are the trace of the induced metric
$\gamma _{ij}$
and extrinsic curvature $K_{\mu \nu }$ on the boundary $\partial \mathcal{M}$%
. To compute the conserved charges of the asymptotically AdS
solutions in Einstein gravity, we use the counterterm approach
\cite{Kraus}. This technique was inspired by $AdS/CFT$
correspondence and consists in adding suitable counterterms
$\mathcal{I}_{ct}$ to the action of the theory in order to ensure
the finiteness of the boundary stress tensor derived by the
quasilocal energy definition \cite{Brown}. Therefore we supplement
the general action (which contains the boundary terms (\ref{Ib}))
with the following boundary counterterm
\begin{equation}
\mathcal{I}_{ct}=-\frac{1}{8\pi }\int_{\partial \mathcal{M}}d^{n}x\sqrt{%
-\gamma }\left( \frac{n-1}{l}\right) .  \label{Ict}
\end{equation}%
Varying the total action ($\mathcal{I}_{tot}=\mathcal{I}+\mathcal{I}_{b}+%
\mathcal{I}_{ct}$) with respect to the induced metric $\gamma
_{ab}$, we find the divergence-free boundary stress-tensor
\begin{equation}
T^{ab}=\frac{K^{ab}-\left( K+\frac{n-1}{l}\right) \gamma
^{ab}}{8\pi }. \label{Tab}
\end{equation}%
Provided the boundary geometry has an isometry generated by a
Killing vector $\mathcal{\xi }^{a}$, a conserved charge
\begin{equation}
\mathcal{Q}_{\mathcal{\xi }}=\int_{\mathcal{B}}d^{n-1}\varphi \sqrt{\sigma }%
T_{ab}n^{a}\mathcal{\xi }^{b},  \label{quasi}
\end{equation}%
can be associated with the boundary $\mathcal{B}$ \cite{Kraus}. In Eq. (\ref%
{quasi}), $\sigma $ is the determinant of the metric $\sigma _{ij}$, and $%
n^{a}$ is the timelike unit normal vector to the boundary
$\mathcal{B}$. In
the context of counterterm method, the limit in which the boundary $\mathcal{%
B}$ becomes infinite ($\mathcal{B}_{\infty }$) is taken, and the
counterterm prescription ensures that the total action and
conserved charges are finite \cite{Kraus}. Physically, this means
that a collection of observers on the
hypersurface whose metric is $\gamma $ all observe the same value of $%
\mathcal{Q}_{\xi }$ provided this hypersurface has an isometry generated by $%
\xi $. In order to ensure a physical interpretation of our future
solutions (the $T_{_{\widehat{0}\widehat{0}}}$ component of the
energy-momentum tensor in the orthonormal frame, should be
positive), we should fix the sign of the coupling constant $\alpha
$ in term of the exponent $s$ in the following
manner%
\begin{equation}
sgn(\alpha )=\left\{
\begin{array}{cc}
(-1)^{1-s}, & s>\frac{1}{2} \\
(-1)^{-s}, & s<\frac{1}{2}%
\end{array}%
\right.   \label{sgn}
\end{equation}

\section{The $(n+1)$-dimensional Charged Rotating Black Branes with Flat
Horizon \label{Rotating Sol}}

Static black hole solutions of nonlinear Maxwell field with
spherical
horizon have been investigated in \cite%
{BHnonmartinez1,BHnonmartinez2,BHnonmartinez3,BHnonmartinez4}, and
here, we
want to generalize it to the rotating solutions. Consider the $(n+1)$%
-dimensional static metric with flat horizon with the following form%
\begin{equation}
ds^{2}=-F(r)dt^{2}+\frac{dr^{2}}{F(r)}+r^{2}\sum\limits_{i=1}^{n-1}d\phi
_{i}^{2},  \label{Metric}
\end{equation}%
we want to generalize this metric (\ref{Metric}) to the rotating
metric with a global rotation. These kinds of rotating solutions
in Einstein gravity have been introduced in Ref.
\cite{Lemos1,Lemos2}. In order to add angular
momentum to the spacetime, we perform the following rotation boost in the $%
t-\phi _{i}$ planes
\begin{equation}
t\mapsto \Xi t-a_{i}\phi _{i},\hspace{0.5cm}\phi _{i}\mapsto \Xi \phi _{i}-%
\frac{a_{i}}{l^{2}}t,  \label{Tr}
\end{equation}%
for $i=1...[n/2]$, where $[x]$ is the integer part of $x$. The
maximum
number of rotation parameters is due to the fact that the rotation group in $%
(n+1)$- dimensions is $SO(n)$ and therefore the number of
independent rotation parameters is $[n/2]$. Thus the
$(n+1)$-dimensional metric of rotating solutions with $\kappa \leq
\lbrack n/2]$ rotation parameters for flat horizon can be written
as
\begin{eqnarray}
ds^{2} &=&-F(r)\left( \Xi dt-{{\sum_{i=1}^{\kappa }}}a_{i}d\phi
_{i}\right)
^{2}+\frac{dr^{2}}{F(r)}  \nonumber \\
&&+\frac{r^{2}}{l^{4}}{{\sum_{i=1}^{\kappa }}}(a_{i}dt-\Xi
l^{2}d\phi
_{i})^{2}  \nonumber \\
&&-\frac{r^{2}}{l^{2}}{\sum_{i<j}^{\kappa }}(a_{i}d\phi
_{j}-a_{j}d\phi _{i})^{2}+r^{2}dX^{2},  \label{met2}
\end{eqnarray}%
where $\Xi =\sqrt{1+\sum_{i}^{k}a_{i}^{2}/l^{2}}$ and $dX^{2}$ is
the Euclidean metric on the $\left( n-\kappa -1\right)
$-dimensional submanifold. Because of the periodic nature of $\phi
_{i}$, the transformation (\ref{Tr}) is not a proper coordinate
transformation on the
entire manifold. Therefore, the metrics (\ref{Metric}) with $k=0$ and (\ref%
{met2}) can be locally mapped into each other but not globally,
and so they are distinct \cite{Stachel}. If we consider Eq.
(\ref{sgn}), the gauge
potential $A_{\mu }$ and the metric function of the rotating metric (\ref%
{met2}) can be written as
\begin{equation}
A_{\mu }=h(r)\left( \Xi \delta _{\mu }^{0}-\delta _{\mu
}^{i}a_{i}\right) \hspace{0.5cm}(\mathrm{no\;sum\;on}\;i),
\label{Amu2}
\end{equation}%
\begin{eqnarray}
F(r) &=&-\frac{2\Lambda r^{2}}{n(n-1)}-{\frac{m}{r^{n-2}}+\Pi
_{1},}
\label{F(r)2} \\
\Pi _{1} &=&\left\{
\begin{array}{cc}
\frac{\left( 2s-1\right) ^{2}\left( \frac{2(2s-n)^{2}q^{2}}{(2s-1)^{2}}%
\right) ^{s}}{(n-1)\left( 2s-n\right) r^{2(ns-3s+1)/(2s-1)}} & 0<s<\frac{1}{2%
},s<0 \\
&  \\
\frac{(-1)^{n+1}2^{n/2}q^{n}\ln r}{r^{n-2}}, & s=\frac{n}{2} \\
&  \\
\frac{-\left( 2s-1\right) ^{2}\left( \frac{2(2s-n)^{2}q^{2}}{(2s-1)^{2}}%
\right) ^{s}}{(n-1)\left( 2s-n\right) r^{2(ns-3s+1)/(2s-1)}}, & \;\mathrm{%
Otherwise}%
\end{array}%
\right.
\end{eqnarray}%
where $m$ is the integration constant which is related to mass
parameter. and $h(r)=-q\ln r$ \ for $s=\frac{n}{2}$ and
$h(r)=-qr^{(2s-n)/(2s-1)}$ otherwise and then for $s=(n+1)/4$, the
electric field $F_{tr}$, is
proportional to $r^{-2}$ in arbitrary dimension \cite%
{HendiRastegar,HendiGBCIM}. In the last relations $q$ is an
integration constant which is related to the electric charge
parameter. It is clear that for $s=0$ and $s=1/2$, the function
$h(r)$ is constant and we do not have any electromagnetic field
and the field equations reduce to uncharged solutions which
discussed in many literatures and we are not interested in it. In
the rest of the paper we assume that $s\neq 0$,$1/2$. The metric
function $F(r)$ , presented here, differ from the linear higher
dimensional Reissner-Nordstr\"{o}m black hole solutions; it is
notable that the electric charge term in the linear case is
proportional to $r^{-2(n-2)}$, but in the presented metric
function, nonlinear case, this term also depends on the exponent
parameter $s$. But, in the linear limit ($s=1$), these solutions
reduce to the higher dimensional Reissner-Nordstr\"{o}m solutions,
as they should be. In addition, it is easy to show that these
solutions reduce to the solutions obtained in Ref.
\cite{HendiRastegar} for $s=(n+1)/4$. Also, it is notable that the
solutions of the rotating metric has the same as static one. But
in the next sections we show that the rotating metric has
different conserved and thermodynamics quantities, and therefore
the new spacetime is different from static one.
\begin{figure}[tbp]
\resizebox{0.5\textwidth}{!}{  \includegraphics{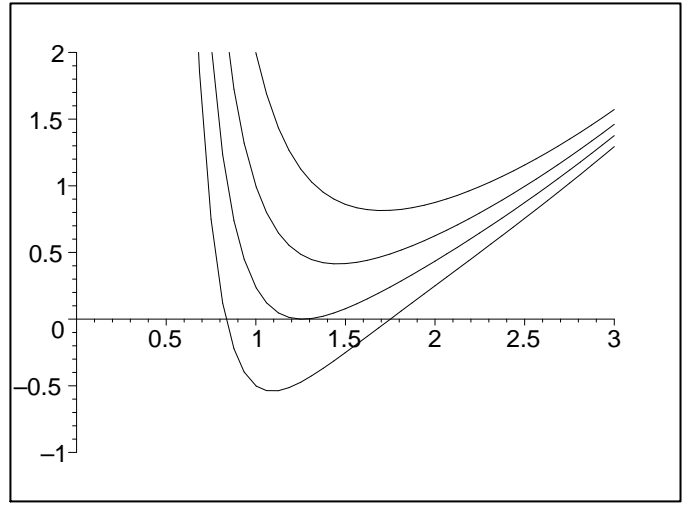}
} 
\caption{$F(r)$ versus $r$ for $n=4$, $q=1$, $s=1$, $\Lambda =-1$, $m<0$, $%
0<m<m_{ext}$, $m=m_{ext}$ and $m>m_{ext}$ from up to down,
respectively.} \label{Fig1F(r)}
\end{figure}
\begin{figure}[tbp]
\resizebox{0.5\textwidth}{!}{  \includegraphics{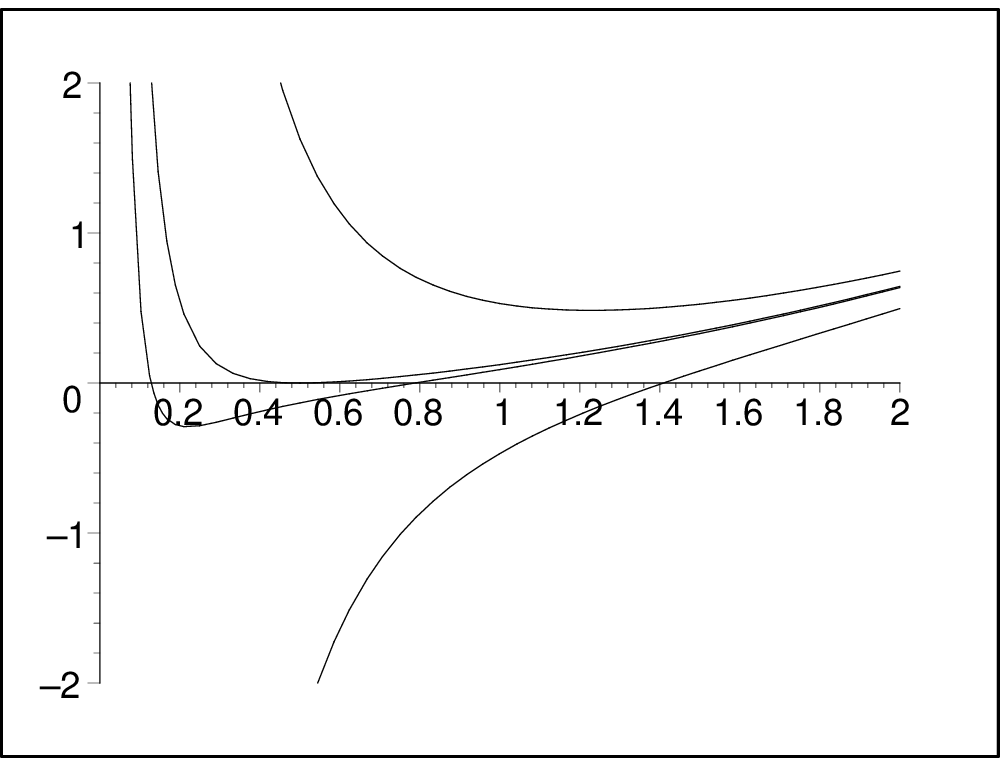}
} 
\caption{$F(r)$ versus $r$ for $n=4$, $q=1$, $s=3$, $\Lambda =-1$, $%
m<m_{ext}<0$, $m=m_{ext}<0$, $m_{ext}<m<0$ and $m>0$ from up to
down, respectively.} \label{Fig2F(r)}
\end{figure}
\begin{figure}[tbp]
\resizebox{0.5\textwidth}{!}{  \includegraphics{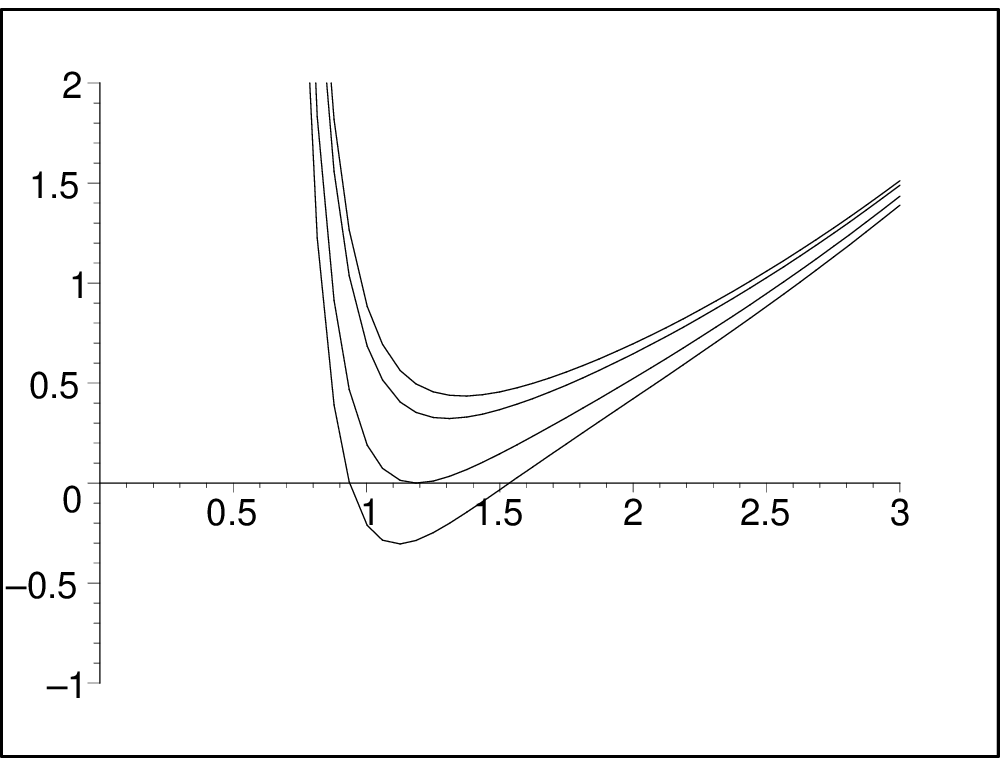}
} 
\caption{$F(r)$ versus $r$ for $n=4$, $q=1$, $s=3/4$, $\Lambda =-1$, $m<0$, $%
0<m<m_{ext}$, $m=m_{ext}$ and $m>m_{ext}$ from up to down,
respectively.} \label{Fig3F(r)}
\end{figure}
\begin{figure}[tbp]
\resizebox{0.5\textwidth}{!}{  \includegraphics{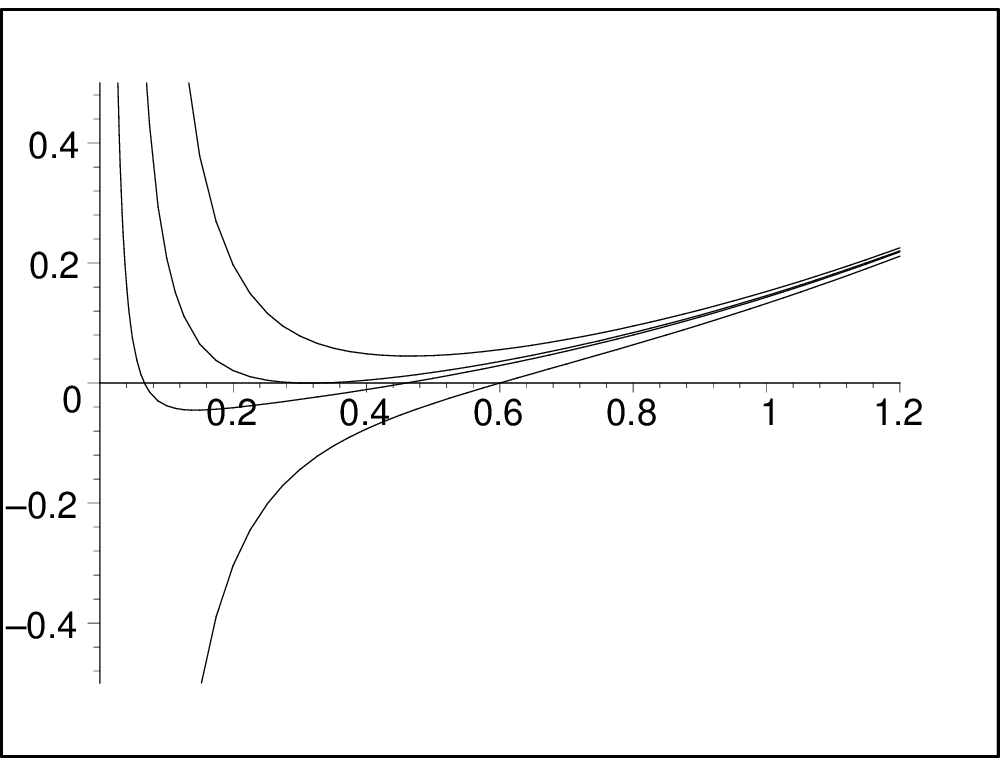}
} 
\caption{$F(r)$ versus $r$ for $n=4$, $q=1$, $s=-3$, $\Lambda =-1$, $%
m<m_{ext}<0$, $m=m_{ext}<0$, $m_{ext}<m<0$ and $m>0$ from up to
down, respectively.} \label{Fig4F(r)}
\end{figure}
\begin{figure}[tbp]
\resizebox{0.5\textwidth}{!}{  \includegraphics{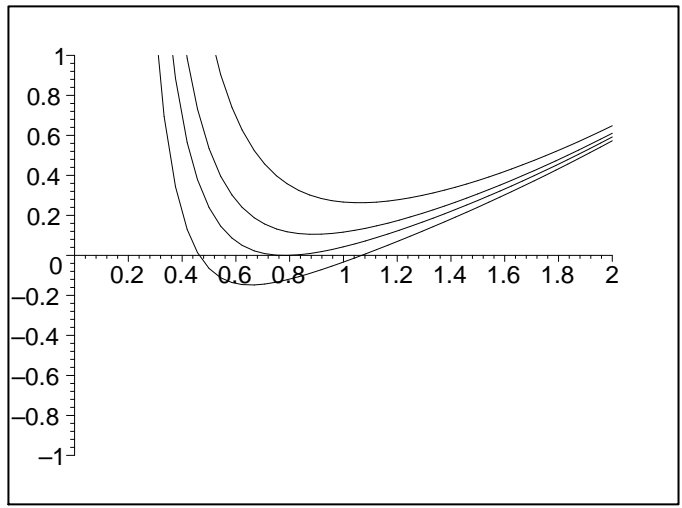}
} 
\caption{$F(r)$ versus $r$ for $n=4$, $q=0.5$, $s=n/2$, $\Lambda
=-1$, $m<0$ , $0<m<m_{ext}$, $m=m_{ext}$ and $m>m_{ext}$ from up
to down, respectively.} \label{Fig5F(r)}
\end{figure}

\subsection{Properties of the solutions}

In order to study the general structure of these spacetime, we
first investigate the effects of the nonlinearity on the
asymptotic behavior of the solutions. It is worthwhile to mention
that for $0<s<\frac{1}{2}$, the asymptotic dominant term of Eq.
(\ref{F(r)2}) is third term and the
presented solutions are not asymptotically AdS, but for the cases $s<0$\ or $%
s>\frac{1}{2}$ (include of $s=\frac{n}{2}$), the asymptotic
behavior of rotating Einstein-nonlinear Maxwell field solutions
are the same as linear AdS case. It is easy to show that the
electromagnetic field is zero for the cases $s=0,\frac{1}{2}$, and
the metric function (\ref{F(r)2}) does not possess a charge term
and it corresponds to uncharged asymptotically AdS one. Second, we
look for the essential singularities. One can show that all
curvature invariants (such as Kretschmann scalar, Ricci square,
Weyl square
and so on) are functions of $F^{\prime \prime }$, $F^{\prime }/r$ and $%
F/r^{2}$ and therefore it is sufficient to study the Kretschmann
scalar for the investigation of the spacetime curvature. After
some algebraic manipulation, we can show that the Kretschmann
scalar diverges at $r=0$ and
is finite for $r\neq 0$. Thus, there is a curvature singularity located at $%
r=0$. The metric (\ref{met2}) has two types of horizons: a Killing
horizon and an event horizon. The Killing horizon is a null
surface whose null generators are tangent to a Killing field. It
is proved that a stationary black hole event horizon should be a
Killing horizon in four-dimensional Einstein gravity
\cite{Hawking1,Hawking2}. In general this proof is not valid for
higher dimensional black holes, but the result is true for all the
known static solutions. The event horizon is the hypersurface in
which light can no longer escape from the gravitational pull of a
black hole. For calculation of the event horizon, one can use the
time dilation interpretation (gravitational red shift)
\cite{Hawking2,Weinberg,MTW}. It is straightforward to show that
the event horizon of the presented rotating solutions are located
at the root(s) of $F(r)=0$ . Numerical calculations (see for e.g.
Figs. (\ref{Fig1F(r)}), (\ref{Fig3F(r)}) and (\ref{Fig5F(r)})
show that for $s=1$ (linear case) and $1/2<s\leq n/2$, the metric of Eqs. (%
\ref{met2}) and (\ref{F(r)2}) has two inner and outer event
horizons located
at $r_{-}$ and $r_{+}$, provided the mass parameter $m$ is greater than $m_{%
\mathrm{ext}}$ given as
\begin{eqnarray}
m_{\mathrm{ext}} &=&\frac{4\Lambda r_{\mathrm{ext}}^{n}}{n(n-1)(n-2)}+\frac{%
\Pi _{2}}{(n-2)} \\
\Pi _{2} &=&\left\{
\begin{array}{cc}
2^{n/2}q_{\mathrm{ext}}^{n}\left[ 1-(n-2)\ln r_{\mathrm{ext}}\right] , & s=%
\frac{n}{2} \\
&  \\
\frac{2(2s-1)(ns-3s+1)\left( \frac{2q_{\mathrm{ext}}^{2}(n-2s)^{2}}{%
(2s-1)^{2}}\right)
^{s}}{(n-1)(n-2s)}r_{\mathrm{ext}}^{\frac{2s-n}{2s-1}}, &
\frac{1}{2}<s<\frac{n}{2}%
\end{array}%
\right.  \nonumber
\end{eqnarray}%
These solutions present naked singularity for $m<m_{\mathrm{ext}}$ and when $%
m=m_{\mathrm{ext}}$, we have an extreme black brane with horizon
radius
\begin{equation}
r_{\mathrm{ext}}=\left\{
\begin{array}{cc}
\left( \frac{-2^{n/2}(n-1)q_{\mathrm{ext}}^{n}}{2\Lambda }\right)
^{1/n}, &
s=\frac{n}{2} \\
\left[ \frac{-(2s-1)}{2\Lambda }\left( \frac{2q_{\mathrm{ext}}^{2}(n-2s)^{2}%
}{(2s-1)^{2}}\right) ^{s}\right] ^{\frac{(2s-1)}{2s(n-1)}}, & \frac{1}{2}<s<%
\frac{n}{2}%
\end{array}%
\right. .
\end{equation}
For $s=0,1/2$, the solutions does not have any charge term and
they are
uncharged black holes, then these cases are excluded from the discussion of $%
m_{\mathrm{ext}}$. Also for $0<s<1/2$, the metric function does
not have a good behavior asymptotically and we are not interested
in this case. Unlike the Reissner-Nordstr\"{o}m solutions ($s=1$)
or $1/2<s\leq n/2$, for $s>n/2$ and $s<0$, the singularity at
$r=0$ for the solutions with non-negative mass
is spacelike, and therefore it is unavoidable. This is due to the fact that $%
f(r)$ approaches to $-\infty $ as $r$ goes to zero and goes to
$+\infty $ at large $r$. These solutions with positive mass
present black branes with one horizon and\ these spacetime with
negative mass presents a naked singularity if
$m<m_{\mathrm{ext}}<0$, an extreme black brane for
$m=m_{\mathrm{ext}}<0$
and a black brane with inner and outer horizons provided $m_{\mathrm{ext}%
}<m<0$, where $m_{\mathrm{ext}}$ is
\begin{eqnarray}
m_{\mathrm{ext}} &=&\frac{4\Lambda
r_{\mathrm{ext}}^{n}}{n(n-1)(n-2)}\mp
\frac{2(2s-1)(ns-3s+1)}{(n-1)(n-2)(n-2s)}\times  \nonumber \\
&&\left( \frac{2q_{\mathrm{ext}}^{2}(n-2s)^{2}}{(2s-1)^{2}}\right) ^{s}r_{%
\mathrm{ext}}^{(2s-n)/(2s-1)},
\end{eqnarray}
where
\[
r_{\mathrm{ext}}=\left[ \frac{\pm (2s-1)}{2\Lambda }\left( \frac{2q_{\mathrm{%
ext}}^{2}(n-2s)^{2}}{(2s-1)^{2}}\right) ^{s}\right]
^{(2s-1)/[2s(n-1)]},
\]%
and $-$ or $+$ correspond to $s>n/2$ or $s<0$, respectively (see figures (%
\ref{Fig2F(r)}) and (\ref{Fig4F(r)}) for clarify).

In some of the classic solutions of the general relativity
equations (e.g. Schwarzschild and Robertson-Walker) no such closed
causal curves occur, although several examples of spacetimes which
do admit closed timelike
curves are known \cite%
{CCC11,CCC12,CCC13,CCC14,CCC21,CCC22,CCC23,CCC24,CCC31,CCC32,CCC33,CCC34,CCC35,CCC36,CCC41,CCC42,CCC43}%
. The possibilities of formation of closed causal curves in
spacetime, within the framework of general relativity have been
discussed in many literatures
\cite{CCC31,CCC32,CCC33,CCC34,CCC35,CCC36}. Also, the rotating
black objects have a net angular momentum, in which its value may
approach a critical value. In this \textquotedblleft
under-rotating\textquotedblright\ case (under critical value) the
closed timelike curves are enclosed inside a velocity of light
surface which lies inside the horizon. In the over-rotating case
there are closed timelike curves outside what superficially
appears to be a horizon \cite{CCC41,CCC42,CCC43}. If we
consider the rotating $(n+1)$-dimensional metric in the following form%
\begin{equation}
ds^{2}=-Vdt^{2}+2Wdtd\phi +Xd\phi ^{2}+g_{ab}dx^{a}dx^{b},
\label{Rotmet}
\end{equation}%
where $a,b=1,2,....,n-1$, and $V$, $X$ and $W$ depend on $x^{a}$.
Ergospheres are given by $V=0$ and the velocity of light surface
is given by $X=0$. Comparing our rotating spacetime (\ref{met2})
with the former metric,
one can show that $X=0$ leads to%
\begin{equation}
F(r)=-\frac{\Xi ^{2}-2}{\Xi ^{2}-1}\frac{r^{2}}{l^{2}},
\end{equation}%
where for sufficiently large rotation parameter(s) ($\Xi
\longrightarrow \infty $), reduces to $F(r)=-\frac{r^{2}}{l^{2}}$.
This result is overlap to our solution for sufficiently large
distance from the black hole and positive cosmological constant
($\Lambda =n(n-1)/2l^{2}$). Thus in the presented spacetime
(\ref{met2}) with negative $\Lambda $, there is not closed
timelike curves.

Although our solution is not static, the Killing vector
\begin{equation}
\chi =\partial _{t}+{\sum_{i=1}^{\kappa }}\Omega _{i}\partial
_{\phi _{i}}, \label{Kil}
\end{equation}%
is the null generator of the event horizon where $\Omega _{i}$ is
the $i$th
component of angular velocity of the outer horizon. The angular velocities $%
\Omega _{i}$'s may be obtained by analytic continuation of the
metric. Setting $a_{i}\rightarrow ia_{i}$ yields the Euclidean
section of (\ref{met2} ), whose regularity at $r=r_{+}$ requires
that we should identify $\phi _{i}\sim \phi _{i}+\beta _{+}\Omega
_{i}$. One obtains
\begin{equation}
\Omega _{i}=\frac{a_{i}}{\Xi l^{2}}  \label{Om}
\end{equation}%
The temperature may be obtained through the use of the definition
of surface gravity,
\[
T_{+}=\beta _{+}^{-1}=\frac{1}{2\pi }\sqrt{-\frac{1}{2}\left(
\nabla _{\mu }\chi _{\nu }\right) \left( \nabla ^{\mu }\chi ^{\nu
}\right) }
\]%
where $\chi $ is the Killing vector (\ref{Kil}). One obtains
\begin{eqnarray}
T_{+} &=&\frac{-\Lambda r_{+}}{2\pi \Xi (n-1)}+\frac{\Pi
_{3}}{4\pi \Xi }
\label{Temp} \\
\Pi _{3} &=&\left\{
\begin{array}{cc}
\frac{(2s-1)\left( \frac{2(n-2s)^{2}q^{2}}{(2s-1)^{2}}\right) ^{s}}{%
(n-1)r_{+}^{(2ns-4s+1)/(2s-1)}} & 0<s<\frac{1}{2},s<0 \\
&  \\
\frac{-2^{n/2}q^{n}}{r_{+}^{n-1}}, & s=\frac{n}{2} \\
&  \\
\frac{-(2s-1)\left( \frac{2(n-2s)^{2}q^{2}}{(2s-1)^{2}}\right) ^{s}}{%
(n-1)r_{+}^{(2ns-4s+1)/(2s-1)}}, & \;\mathrm{Otherwise}%
\end{array}%
\right. .
\end{eqnarray}

\subsection{Conserved quantities \label{Therm}}

More than thirty years ago, Bekenstein argued that the entropy of
a black hole is a linear function of the area of its event
horizon, which so-called area law \cite{Bekenstein1,Bekenstein2}.
He also proposed a value for the proportionality constant, deduced
from a semiclassical calculation of the minimum increase in the
area of a black hole when it absorbs a particle. Bekenstein's
statements can in fact be generalized by considering the quantum
nature of the particle and taking then into account the
uncertainty principle and the energy-momentum dispersion relation
\cite{Hod}. This generalized argument leads essentially to the
same conclusion about the linearity of the entropy with respect to
the black hole area. Since the area law of the entropy is
universal, and applies to all kinds of black holes/branes in
Einstein gravity \cite{Bekenstein1,Bekenstein2,Hawking3},
therefore the entropy per unit volume $V_{n-1}$ of the presented
black branes is equal to one-quarter of the area of the horizon,
i.e.,
\begin{equation}
S=\frac{\Xi }{4}r_{+}^{n-1}.  \label{Entropy}
\end{equation}%
The electric charge per unit volume $V_{n-1}$ of the black branes,
$Q$, can be found by calculating the flux of the electromagnetic
field at infinity, yielding
\begin{equation}
Q=\left\{
\begin{array}{cc}
\frac{(-1)^{(n+2)/2}2^{n/2}\Xi q^{n-1}}{8\pi }, & s=\frac{n}{2} \\
&  \\
\frac{(-1)^{s+1}2^{s}\Xi }{8\pi }\left(
\frac{(2s-n)q}{(2s-1)}\right)
^{2s-1}, & \;\mathrm{otherwise}%
\end{array}%
\right. .  \label{Charg}
\end{equation}%
The present spacetime (\ref{met2}), have boundaries with timelike
($\xi =\partial /\partial t$) and rotational ($\varsigma =\partial
/\partial \varphi $) Killing vector fields. From Eq.
(\ref{quasi}), one obtains the quasilocal mass and angular
momentum
\begin{eqnarray}
M &=&\int_{\mathcal{B}}d^{n-1}\varphi \sqrt{\sigma }T_{ab}n^{a}\xi ^{b}=%
\frac{V_{n-1}}{16\pi }\left( n\Xi ^{2}-1\right) m  \label{Mas} \\
J &=&\int_{\mathcal{B}}d^{n-1}\varphi \sqrt{\sigma
}T_{ab}n^{a}\varsigma ^{b}=\frac{V_{n-1}}{16\pi }n\Xi ma_{i}
\label{Amom}
\end{eqnarray}%
provided the hypersurface $\mathcal{B}$ contains the orbits of
$\varsigma $.
Eqs. (\ref{Charg})-(\ref{Amom}) show that, like the Born-Infeld gravity \cite%
{BlackBrane1,BlackBrane2,BlackBrane3,Higher1,Higher2,Higher3,Higher4,Higher5}%
, although, the presence of the nonlinear Maxwell field has no
effects on the mass and angular momentum of the black brane, the
nonlinear parameter, $s $, change the electrical charge of black
brane. It is a matter of straightforward calculation to show that
the conserved and thermodynamics quantities satisfy the first law
of thermodynamics
\begin{equation}
dM=TdS+{{{\sum_{i=1}^{k}}}}\Omega _{i}dJ_{i}+\Phi dQ.
\label{FirstLaw}
\end{equation}%
It is notable that, in general, the temperature and the other
conserved
quantities are independent of the choice of coordinates. Eqs. (\ref{Temp})-(%
\ref{Amom}) show that the thermodynamics and conserved quantities
are different for static and rotating spacetimes, and
consequently, confirm that the static and rotating metrics are
distinct and they describe two different spacetimes.

\section{CLOSING REMARKS}

In this paper we generalized the solutions of Ref.
\cite{HendiRastegar} for arbitrary values of nonlinear parameter
$s$. At first, we considered Einstein-Hilbert action with
cosmological constant coupled to an abelian gauge field for which
the Lagrangian is given by a power of the Maxwell invariant. The
presented nonlinear parameter effected on the electromagnetic
field and charge term of metric functions. One can show that these
solutions reduce to the solutions of linear and conformally
invariant Maxwell field provided the nonlinear parameter $s=1$ and
$s=(n+1)/4 $, respectively.

Then, by a suitable rotation boost, we presented the horizon flat
rotating spacetime with $\kappa \leq \lbrack n/2]$ rotation
parameters. We investigated the asymptotic behavior of these
solutions and found that for the special values of the nonlinear
parameter $s$, the solutions are not asymptotically AdS. Also, we
found the essential singularity located at $r=0$ and showed that
for a suitable values of $s$, the presented solutions may be
interpreted as black brane solutions with two event horizons,
extreme black hole or naked singularity provided the mass
parameter, $m$, of the solutions is more than, less than or equal
to the extreme value $m_{ext}$.

In addition, it is worth to mention that for $s>n/2$ and $s<0$,
the singularity at $r=0$ for the solutions with non-negative mass
is spacelike. This is due to the fact that $f(r)$ approaches to
$-\infty $ as $r$ goes to zero and goes to $+\infty $ at large
$r$. These solutions with positive mass present black holes with
one horizon. Also, we investigated the existence of closed
timelike curves and showed that they can exist for sufficiently
large distance from the black hole and sufficiently large rotation
parameter, only for positive cosmological constant.

At last, using the counterterm approach, area and Gauss laws, we
computed the quasi local mass and angular momentum, entropy and
electrical charge of the rotating black branes. These calculations
showed that only the electrical charge depend on the nonlinear
parameter. In fact the nonlinearity effects on the electric field,
the behavior of spacetime, temperature, type of singularity,
electrical charge and all the expressions include of the charge
parameter. Finally, it is worthwhile to examine the
thermodynamical stability as well as dynamical (gravitational)
stability of these black branes.

This work has been supported financially by Research Institute for
Astronomy and Astrophysics of Maragha.


\begin{thebibliography}{99}
\bibitem{BI1} M. Born and L. Infeld, Proc. R. Soc. London A \textbf{143},
410 (1934)

\bibitem{BI2} M. Born and L. Infeld, Proc. R. Soc. London A \textbf{144},
425 (1934)

\bibitem{Plebanski} J. Plebanski, \textit{Lectures on Non--Linear
Electrodynamics} (Nordita, 1968)

\bibitem{FradTsey} E. S. Fradkin and A. Tseytlin, Phys. Lett. B, \textbf{163}%
, 123 (1985)

\bibitem{Tsey} A. Tseytlin, Nucl. Phys B \textbf{276}, 391 (1986)

\bibitem{SeibWitt} N. Seiberg and E. Witten, JHEP \textbf{09}, 032 (1999)

\bibitem{Rasheed} D. A. Rasheed, [arXiv:9702087]

\bibitem{Novello11} M. Novello and S. E. Perez Bergliaffa, Phys. Rept.
\textbf{463}, 127 (2008)

\bibitem{Novello12} M. Novello, E. Goulart, J. M. Salim and S. E. Perez
Bergliaffa, Class. Quant. Grav. \textbf{24}, 3021 (2007)

\bibitem{Novello13} M. Novello, S. E. Perez Bergliaffa and J. Salim, Phys.
Rev. D \textbf{69}, 127301 (2004)

\bibitem{Novello14} V. A. De Lorenci, R. Klippert, M. Novello and J. M.
Salim, Phys. Rev. D \textbf{65}, 063501 (2002)

\bibitem{Novello2} M. Novello, Aline N. Araujo and J. M. Salim,
[arXiv:0802.1875]

\bibitem{Camara1} C.S. Camara, J. C. Carvalho and M. R. De Garcia Maia, Int.
J. Mod. Phys. D \textbf{16}, 427 (2007)

\bibitem{Camara2} V. V. Dyadichev, D. V. Gal'tsov and P. V. Moniz, AIP Conf.
Proc. \textbf{861}, 312 (2006)

\bibitem{Camara3} V. V. Dyadichev, D. V. Gal'tsov and P. V. Moniz, Phys.
Rev. D \textbf{72}, 084021 (2005)

\bibitem{Camara4} M. Novello, Int. J. Mod. Phys. A \textbf{20}, 2421 (2005)

\bibitem{Camara5} M. Novello, AIP Conf. Proc. \textbf{782}, 306 (2005)

\bibitem{Camara6} R. Garcia-Salcedo and N. Breton, Class. Quant. Grav.
\textbf{22}, 4783 (2005)

\bibitem{Camara7} C. S. Camara, M. R. de Garcia Maia, J. C. Carvalho and J.
A. S. Lima, Phys. Rev. D \textbf{69}, 123504 (2004)

\bibitem{Camara8} R. Garcia-Salcedo and N. Breton, Class. Quant. Grav.
\textbf{20}, 5425 (2003)

\bibitem{Camara9} E. Elizalde, J. E. Lidsey, S. Nojiri and S. D. Odintsov,
Phys. Lett. B \textbf{574}, 1 (2003)

\bibitem{Camara10} D. N. Vollick, Gen. Rel. Grav. \textbf{35}, 1511 (2003)

\bibitem{Camara11} V. V. Dyadichev, D. V. Gal'tsov, A. G. Zorin and M. Yu.
Zotov, Phys. Rev. D \textbf{65}, 084007 (2002)

\bibitem{Camara12} P. V. Moniz, Phys. Rev. D \textbf{66}, 103501 (2002)

\bibitem{Camara13} P. Moniz, Class. Quant. Grav. \textbf{19}, L127 (2002)

\bibitem{Camara14} R. Garcia-Salcedo and N. Breton, Int. J. Mod. Phys. A
\textbf{15}, 4341 (2000)

\bibitem{Camara15} B. L. Altshuler, Class. Quant. Grav. \textbf{7}, 189
(1990)

\bibitem{Vollick} D. N. Vollick, Phys. Rev. D \textbf{78}, 063524 (2008)

\bibitem{AhGubMald} O. Aharony, S. S. Gubser, J. Maldacena, H. Ooguri and Y.
Oz, Phys. Rept. \textbf{323}, 183 (2000)

\bibitem{Callan} C. G. Callan, A. Guijosa, K. G. Savvidy and O. Tafjord,
Nucl. Phys B \textbf{555}, 183 (1999)

\bibitem{Vargas1} P. Vargas Moniz, Phys. Rev. D \textbf{66}, 103501 (2002)

\bibitem{Vargas2} P. Vargas Moniz, Class. Quant. Grav. \textbf{19}, L127
(2002)

\bibitem{Garcia1} E. Ayon-Beato and A. Garcia, Phys. Rev. Lett. \textbf{80},
5056 (1998)

\bibitem{Garcia2} E. Ayon-Beato and A. Garcia, Phys. Lett. B \textbf{464},
25 (1999)

\bibitem{Garcia3} E. Ayon-Beato and A. Garcia, Gen. Rel. Grav. \textbf{31},
629 (1999)

\bibitem{Garcia4} E. Ayon-Beato and A. Garcia, Gen. Rel.Grav. \textbf{37},
635 (2005)

\bibitem{Breton} N. Breton, Phys. Rev. D \textbf{67}, 124004 (2003)

\bibitem{BHnon1} R. G. Cai, D. W. Pang and A. Wang, Phys. Rev. D \textbf{70}%
, 124034 (2004)

\bibitem{BHnon2} R. G. Cai and Y. W. Sun, JHEP \textbf{09}, 115 (2008)

\bibitem{BHnon3} S. S. Yazadjiev, Phys. Rev. D \textbf{72}, 044006 (2005)

\bibitem{BHnon4} D. N. Vollick, Phys. Rev. D \textbf{72}, 084026 (2005)

\bibitem{BHnon5} Y. S. Myung, Y. W. Kim and Y. J. Park, Phys. Rev. D \textbf{%
78}, 044020 (2008)

\bibitem{BHnon6} Y. S. Myung, Y. W. Kim and Y. J. Park, Phys. Rev. D \textbf{%
78}, 084002 (2008)

\bibitem{BHnon7} S. H. Mazharimousavi, M. Halilsoy and Z. Amirabi, Phys.
Rev. D \textbf{78}, 064050 (2008)

\bibitem{BHnon8} A. Khodam-Mohammadi, Grav. Cosmol. \textbf{15}, 154 (2009)

\bibitem{BHnonmartinez1} M. Hassaine and C. Martinez, Phys. Rev. D \textbf{75%
}, 027502 (2007)

\bibitem{BHnonmartinez2} M. Hassaine and C. Martinez, Class. Quant. Grav.
\textbf{25}, 195023 (2008)

\bibitem{BHnonmartinez3} H. Maeda, M. Hassaine and C. Martinez, Phys. Rev. D
\textbf{79}, 044012 (2009)

\bibitem{BHnonmartinez4} S. H. Mazharimousavi and M. Halilsoy, Phys. Lett. B
\textbf{681}, 190 (2009)

\bibitem{BlackBrane1} M. H. Dehghani and H. R. Rastegar Sedehi, Phys. Rev. D
\textbf{74}, 124018 (2006)

\bibitem{BlackBrane2} M. H. Dehghani, S. H. Hendi, A. Sheykhi and H.
Rastegar Sedehi, JCAP \textbf{02}, 020 (2007)

\bibitem{BlackBrane3} S. H. Hendi, J. Math. Phys. \textbf{49}, 082501 (2008)

\bibitem{Bronnikov} K.A. Bronnikov, Phys. Rev. D \textbf{63}, 044005 (2001)

\bibitem{MagBrane1} M. H. Dehghani, N. Bostani and S. H. Hendi, Phys. Rev. D
\textbf{78}, 064031 (2008)

\bibitem{MagBrane2} M. H. Dehghani, A. Sheykhi and S. H. Hendi, Phys. Lett.
B \textbf{659}, 476 (2008 )

\bibitem{Higher1} M. H. Dehghani and S. H. Hendi, Int. J. Mod. Phys. D
\textbf{16}, 1829 (2007)

\bibitem{Higher2} S. Mukherji and S. Pal, [arXiv:08062507]

\bibitem{Higher3} M. H. Dehghani and S. H. Hendi, Gen. Rel. Grav. \textbf{41}%
, 1853 (2009)

\bibitem{Higher4} M. Aiello, R. Ferraro and G. Giribet, Phys. Rev. D \textbf{%
70}, 104014 (2004)

\bibitem{Higher5} M. H. Dehghani, N. Alinejadi and S. H. Hendi, Phys. Rev. D
\textbf{77}, 104025 (2008)

\bibitem{Myers1} R. C. Myers, Phys. Rev. D \textbf{36}, 392 (1987)

\bibitem{Myers2} S. C. Davis, Phys. Rev. D \textbf{67}, 024030 (2003)

\bibitem{Kraus} V. Balasubramanian and P. Kraus, Commun. Math. Phys. \textbf{%
208} (1999) 413

\bibitem{Brown} J. D. Brown and J. W. York, Phys. Rev. D \textbf{47} (1993)
1407

\bibitem{Lemos1} J. P. S. Lemos and V. T. Zanchin, Phys. Rev. D \textbf{54},
3840 (1996)

\bibitem{Lemos2} J. P. S. Lemos, Phys. Lett. B \textbf{353}, 46 (1995)

\bibitem{Stachel} J. Stachel, Phys. Rev. D \textbf{26}, 1281 (1982)

\bibitem{HendiRastegar} S. H. Hendi and H. R. Rastegar-Sedehi, Gen. Rel.
Grav., \textbf{41}, 1355 (2009)

\bibitem{HendiGBCIM} S. H. Hendi, Phys. Lett. B \textbf{677}, 123 (2009)

\bibitem{Hawking1} S.W. Hawking, Commun. Math. Phys. \textbf{25}, 152 (1972)

\bibitem{Hawking2} S.W. Hawking and G. F. R. Ellis, \textit{The Large Scale
Structure of Spacetime} (Cambridge University Press, Cambridge,
England, 1973)

\bibitem{Weinberg} S. Weinberg, \textit{Gravitation and Cosmology} (John
Wiley \& Sons, Inc. New York, 1972)

\bibitem{MTW} C. W. Misner, K. S. Thorne and J. A Wheeler, \textit{%
Gravitation} (W. H. Freeman and Company, San Frencisco, USA, 1972)

\bibitem{CCC11} K. Godel, Rev. Mod. Phys. \textbf{21}, 447 (1949)

\bibitem{CCC12} F. J. Tipler, Phys. Rev. D \textbf{9}, 2203 (1974)

\bibitem{CCC13} H. Monroe, [arXiv:gr-qc/0607134]

\bibitem{CCC14} J. Magueijo and A. Mozaffari, [arXiv:gr-qc/09113697]

\bibitem{CCC21} M. S. Morris, K. S. Thorne and U. Yurtsever, Phys. Rev.
Lett. \textbf{61}, 1446 (1988)

\bibitem{CCC22} J. R. Gott, Phys. Rev. Lett. \textbf{66}, 1126 (1991)

\bibitem{CCC23} O. Bertolami and F. S. N. Lobo, [arXiv:gr-qc/09020559]

\bibitem{CCC24} F. S. N. Lobo, [arXiv:gr-qc/07100428]

\bibitem{CCC31} A. Ori, Phys. Rev. Lett. \textbf{71}, 2517 (1993)

\bibitem{CCC32} A. Ori and Y. Soen, Phys. Rev. D \textbf{49}, 3990 (1994)

\bibitem{CCC33} Y. Soen and A. Ori, Phys. Rev. D \textbf{54}, 4858 (1996)

\bibitem{CCC34} A. Ori, Phys. Rev. Lett. \textbf{95}, 021101 (2005)

\bibitem{CCC35} R. L. Mallett, Found. Phys. \textbf{33}, 1307 (2003)

\bibitem{CCC36} A. Ori, [arXiv:gr-qc/0701024]

\bibitem{CCC41} G. W. Gibbons and C. A. R. Herdeiro, Class. Quant. Grav.
\textbf{16}, 3619 (1999)

\bibitem{CCC42} L. Dyson, [arXiv:hep-th/0608137]

\bibitem{CCC43} E. G. Gimonab and P. Horava, [arXiv:hep-th/0405019]

\bibitem{Bekenstein1} J.D. Bekenstein, Lett. Nuovo Cimento \textbf{4}, 737
(1972)

\bibitem{Bekenstein2} J.D. Bekenstein, Phys. Rev. D \textbf{7}, 2333 (1973)

\bibitem{Hod} S. Hod, Phys. Rev. D \textbf{59}, 024014 (1999).

\bibitem{Hawking31} S. W. Hawking and C. J. Hunter, Phys. Rev. D \textbf{59},
044025 (1999)

\bibitem{Hawking32} S. W. Hawking, C. J. Hunter and D. N. Page, Phys. Rev. D
\textbf{59}, 044033 (1999)

\bibitem{Hawking33} R. B. Mann, Phys. Rev. D \textbf{60}, 104047 (1999)

\bibitem{Hawking34} R. B. Mann, Phys. Rev. D \textbf{61}, 084013 (2000)

\bibitem{Hawking35} C. J. Hunter, Phys. Rev. D \textbf{59}, 024009 (1999)
\end{thebibliography}
\end{document}